\documentclass[%
 reprint,
 amsmath,amssymb,
 aps,
]{revtex4-1}

\usepackage{graphicx}
\usepackage{dcolumn}
\usepackage{bm}
\usepackage{amsmath}
\usepackage{amssymb}
\usepackage{latexsym}
\usepackage{epsfig}
\usepackage{amsbsy}
\usepackage{array}
\usepackage{amssymb}
\usepackage{setspace}
\usepackage{bm}
\usepackage{diagbox}

\def\sint{\ifmmode{- \!\!\!\!\!\! \int}
    \else{\hbox{$- \!\!\!\! \int \ $}}\fi}

\begin{document}


\title{Origin of lobe and cleft at the gravity current front}

\author{C.Y. Xie}
\author{J.J. Tao}
\email{jjtao@pku.edu.cn}
\author{L.S. Zhang}
\affiliation{CAPT-HEDPS, SKLTCS, Collaborative Innovation Center of IFSA, Department of Mechanics Engineering
science, College of Engineering, Peking University, Beijing, 100871 P. R. China}


\begin{abstract}
In this paper the Rayleigh-Taylor instability (RTI) along the density interfaces of gravity-current fronts is analyzed. Both the location and the spanwise wavenumber of the most unstable mode determined by the local dispersion relation agree with those of the strongest perturbation obtained from numerical simulations, suggesting that the original formation mechanism of lobes and clefts at the current front is RTI. Furthermore, the predictions of the semi-infinite RTI model, i.e. the original dominating spanwise wavenumber of the Boussinesq current substantially depends on the Prandtl number and has a 1/3 scaling law with the Grashof number, are confirmed by the three-dimensional numerical simulations.
\end{abstract}



\maketitle

\section{Introduction}
Lobes and clefts are the primary characteristics formed at the fronts of gravity currents \cite{Simpson1982,Simpson1997,Huppert2006,Meiburg2010,Constantinescu2014,Hughes2016}, dominating the local mass and momentum transportation \cite{Cantero2007,Espath15,Jackson13}. Experimental study shown that these three-dimensional structures originated from the instability produced by the overrunning of light fluid by the dense current \cite{Simpson1972}. However, it was argued recently that the overturning of ambient fluid or the density stratification might not be necessary and there might be some other Reynolds number-dependent formation mechanisms for lobes and clefts \cite{McElwaine04}. Linear stability analyses were carried out based on the flow fields of the current fronts obtained numerically, illustrating that the fronts were unstable to three-dimensional modes, and the mode growth rate was a weak function of the Schmidt number $Sc$ unless $Sc$ was very small \cite{Hartel2000b}. Three-dimensional numerical simulations were also carried out, and it was concluded that the lobe and cleft patterns were independent of $Sc$  by comparing the mean spanwise wavelength of lobes obtained numerically and the wavelength of the most unstable mode \cite{Bonometti2008}. However, it should be noted that the mean spanwise wavelengths were estimated by counting the number of lobes observed in experiments and simulations \cite{Hartel2000b,Bonometti2008}, where the nonlinear cleft merging and lobe growing  have occurred \cite{Simpson1972,Cantero2007, Espath15}, and hence the estimated wavelengths don't correspond strictly to the linearly unstable mode.

Since the stratification is influenced by the diffusion property (strong diffusion corresponds to weak stratification), which is represented by the Schmidt number for mass diffusion or the Prandtl number (Pr) for thermal diffusion, a stratification related phenomenon should depend on $Sc$ or $Pr$. In fact, the discrepancy of the spanwise wavenumber between the linear result for $Sc=1$ and the experimental data for $Sc$ of $O(10^3)$ \cite{Hartel2000b} has implied $Sc$ effect. In addition, there are plenty of vortical structures in the viscous flow field of the current front \cite{Simpson1972,Hartel2000a,Peng2010}, and the previous linear stability analyses required detailed information of both the velocity field and the fluid's properties \cite{Hartel2000b}. However, does the velocity field make much contribution to determining the spanwise wavenumber of the most unstable mode? It is shown in this paper by stability analyses and numerical simulations that the original spanwise wavenumber of lobes and clefts is intrinsically determined by the Rayleigh-Taylor instability, which depends on the fluid properties but not the velocity field, and its relations with the Grashof number and the Prantl number are obtained explicitly.

\section{Simulations and stability analyses }
\begin{figure}
\centering
\begin{minipage}[b]{0.50\textwidth}
\centering
\includegraphics[width=1.7in]{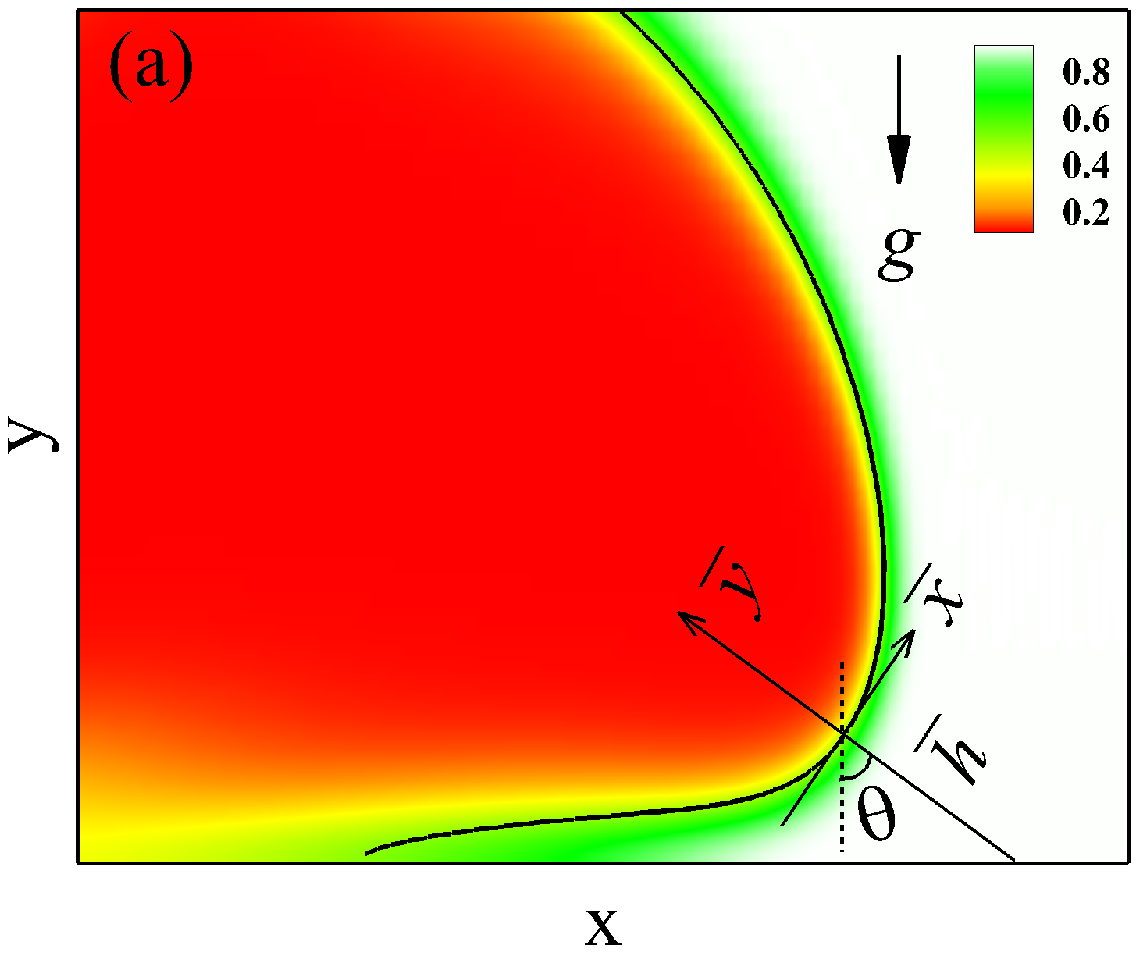}
\includegraphics[width=1.7in]{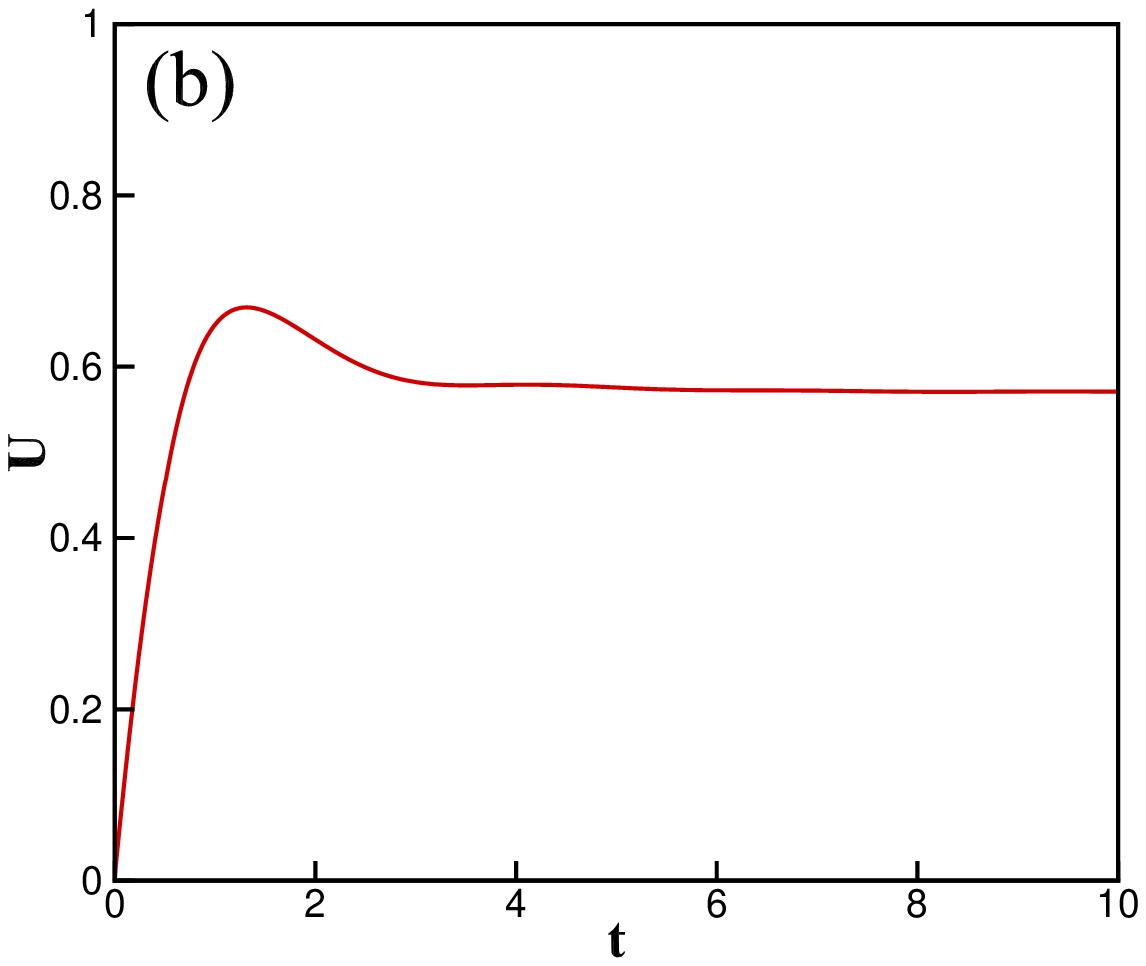}
\end{minipage}
\caption{(a) The dimensionless temperature $T$ at the current front. The solid curve represents $T=0.5$. (b) The front velocity, the velocity of the forefront with $T=0.5$ in the $x$ direction, as a function of time for Pr=1 and Gr=$10^6$ without introducing artificial disturbances. \label{fig1}}
\end{figure}

We consider gravity currents propagating along horizontal adiabatic bottom walls, which are governed by the incompressible Navier-Stokes equations. The density difference is caused by the temperature difference between the current fluid and the ambient fluid $\Delta \Theta$, which is assumed to be small enough to use the Boussinesq approximation. The Grashof number is defined as Gr$=\frac{g\beta\Delta\Theta h^3}{\nu^2}=\frac{2gAh^3}{\nu^2}$, where $g$, $\beta$, $\nu$, $h$ and $A$ are the gravitational acceleration, the thermal expansion coefficient, the kinematic viscosity, the characteristic length scale, and the Atwood number, respectively. The diffusion ratio is described by the Prandtl number, which is equivalent to the Schmidt number and is defined as Pr$=\nu/\kappa$, where $\kappa$ is the thermal diffusivity. By using the characteristic temperature $\Delta\Theta$, velocity $\nu \sqrt{Gr}/h$ and length $h$, the dimensionless governing equations are obtained as
\begin{eqnarray}
\label{eq:eq-1}
 && \partial_{t}\mathbf{u}+(\mathbf{u}\cdot\nabla)\mathbf{u}=-\nabla p+\frac{1}{\sqrt{Gr}}\nabla^2\mathbf{u}+T\mathbf{e_y},\\
 && \partial_{t}T+(\mathbf{u}\cdot\nabla)T=\frac{1}{\sqrt{Gr}Pr}\nabla^2 T,\\
 && \nabla\cdot\mathbf{u}=0,
\end{eqnarray}
where $\mathbf{e_y}$ is the unit vector contrary to the gravity direction. These equations are solved with a pseudo-spectral method \cite{Chevalier2007} to study the lock-exchange flows in a cuboid box of $(L_x,L_y,L_z)$ with periodic boundary conditions in the streamwise direction $x$ and the spanwise direction $z$, and no-slip conditions in the vertical or transverse direction $y$. The flow field is expanded in Fourier modes in the $x$ and $z$ directions and Chebyshev polynomials in $y$ direction.  The fluids are initially at rest and the dense fluid fills two ends of the domain, where $0\le x\le L_x/4$ and $3L_x/4\le x\le L_x$.

\begin{table}[!htbp]
\small
\centering  \caption{The domain size $(L_x, L_z)$  used in the simulations}\label{Tab1}
\begin{tabular}{|l|c|c|c|c|c|}
\hline
\diagbox{Pr}{Gr} &$10^4$            &$10^5$&$10^6$&$10^7$&$10^8$\\ 
\hline
$0.2$  &(48, 83.1) &(48, 42.4)     &(48, 19.6)     &(48, 9.1)     &(48, 4.2)\\
\hline
$1$    &(48, 43.4) &(48, 19.9)     &(48, 10.0)       &(48, 4.4)     &(48, 2.0)\\
\hline
$5$    &(48, 28.5) &(48, 12.7)     &(48, 5.9)     &(48, 2.8)     &(24, 1.3) \\
\hline
$650$  &(48, 21.0) &(48, 9.4)      &(48, 4.3)     &(24, 2.0)     & (12, 0.9) \\
\hline
\end{tabular}
\end{table}

After a short acceleration phase, the gravity current develops and reaches its slumping phase \cite{Benjamin1968,Rottman1983,Klemp1994,Shin2004,Scotti2008,Borden2013}, where the front speed maintains constant. In this paper, we will focus on the three-dimensional instability at the beginning of the slumping phase. In order to examine the original dominating wavenumber $k_{max}$, the spanwise wavenumber of the linear unstable mode with the largest growth rate, random temperature disturbances are added to the whole field near the beginning of the slumping phase with the form of $\theta '(x,y,z)=\delta \Sigma_{n=1}^{n_{dist}}\cos [2\pi nz/L_z+\phi (n)]$, where the amplitude $\delta$ should be small in order to trigger the linearly unstable mode and is set as 0.00033, $n_{dist}$ is the disturbing mode number, and $\phi(n)$ is a random number in the range of $(0,2\pi]$. For each couple of $Pr$ and $Gr$, preliminary simulations were carried out first to obtain the approximate dominating spanwise wavelength of lobes and clefts, and then $L_z$ is chosen to be at least twenty times larger than the approximation. The method determining the dominating wavelength or wavenumber is introduced next.

 In the present simulations, $L_y=2$, spectral modes $(n_x, n_y, n_z)=(2048, 129, 1024)$, and ($L_x, L_z$) used for different parameters are listed in Table \ref{Tab1}.  Both the domain width and the mode resolution are larger and higher than those used in the previous three-dimensional simulations (e.g. \cite{Bonometti2008}) respectively, and it was checked that the dominating wavenumbers obtained with half modes agreed with the present results to better than $3\%$. Since vortex rolls will be induced by Kelvin-Helmholtz instability at high Grs and Prs \cite{Hartel2000b}, shorter $L_x$ and then finer resolutions are used as shown in Table \ref{Tab1}. The front velocities and the dominating wavenumbers obtained from the test simulations within domains of $3L_x/4$ were checked to be consistent with the present results, indicating that $L_x$ is long enough to study the spanwise instability at the beginning stage of the slumping phase. In addition, the slumping velocity shown in Fig.1(b) and the dominating spanwise wavenumbers shown in Fig. 4(a) for $Pr=1$ agree well with the previous data (Fig. 3 and Fig. 6 of \cite{Hartel2000b}) respectively, confirming that the present computational configurations are applicable.

\begin{figure}
\centering
\begin{minipage}[b]{0.50\textwidth}
\centering
\includegraphics[width=3.60in]{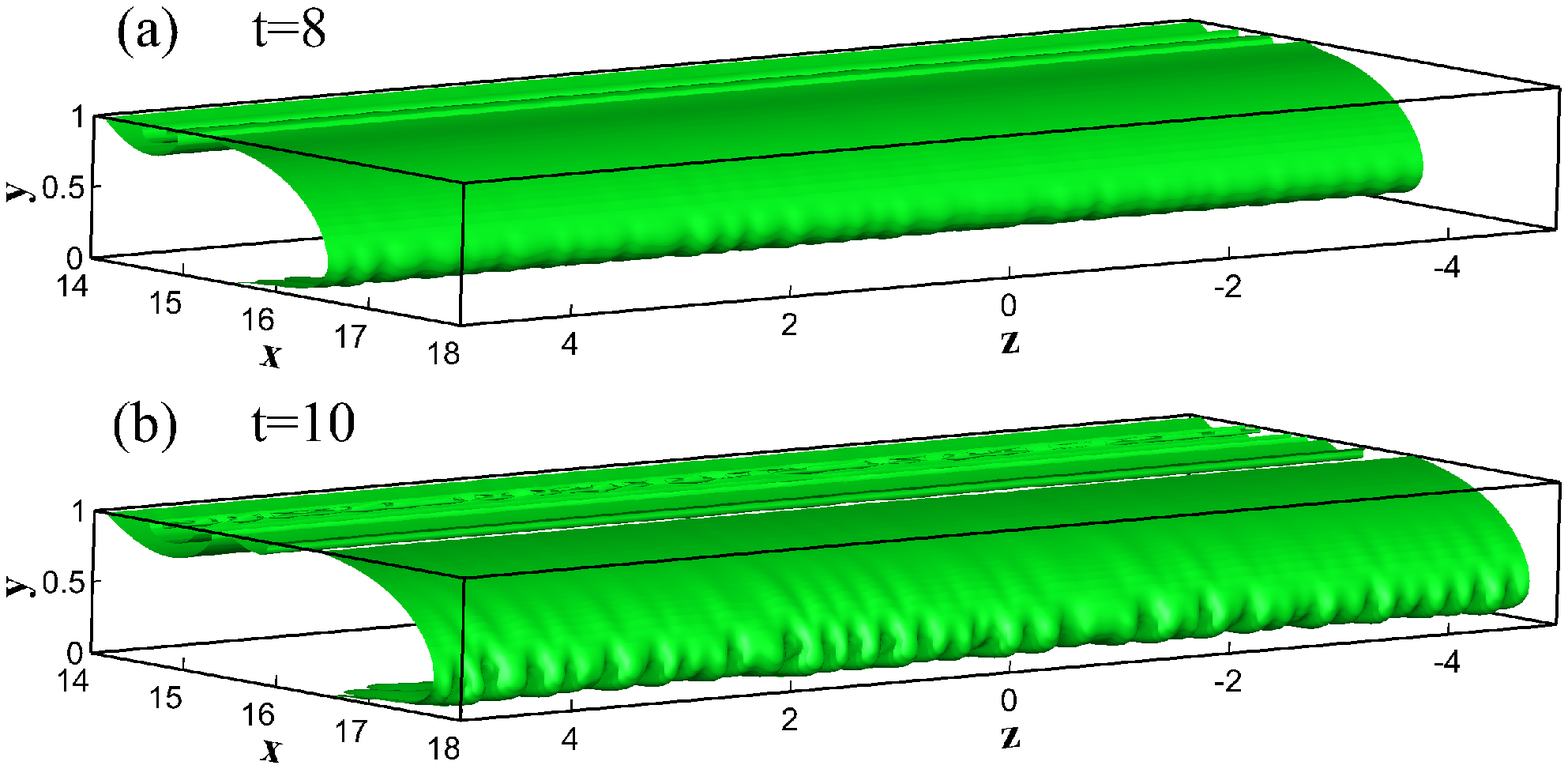}
\includegraphics[width=1.68in]{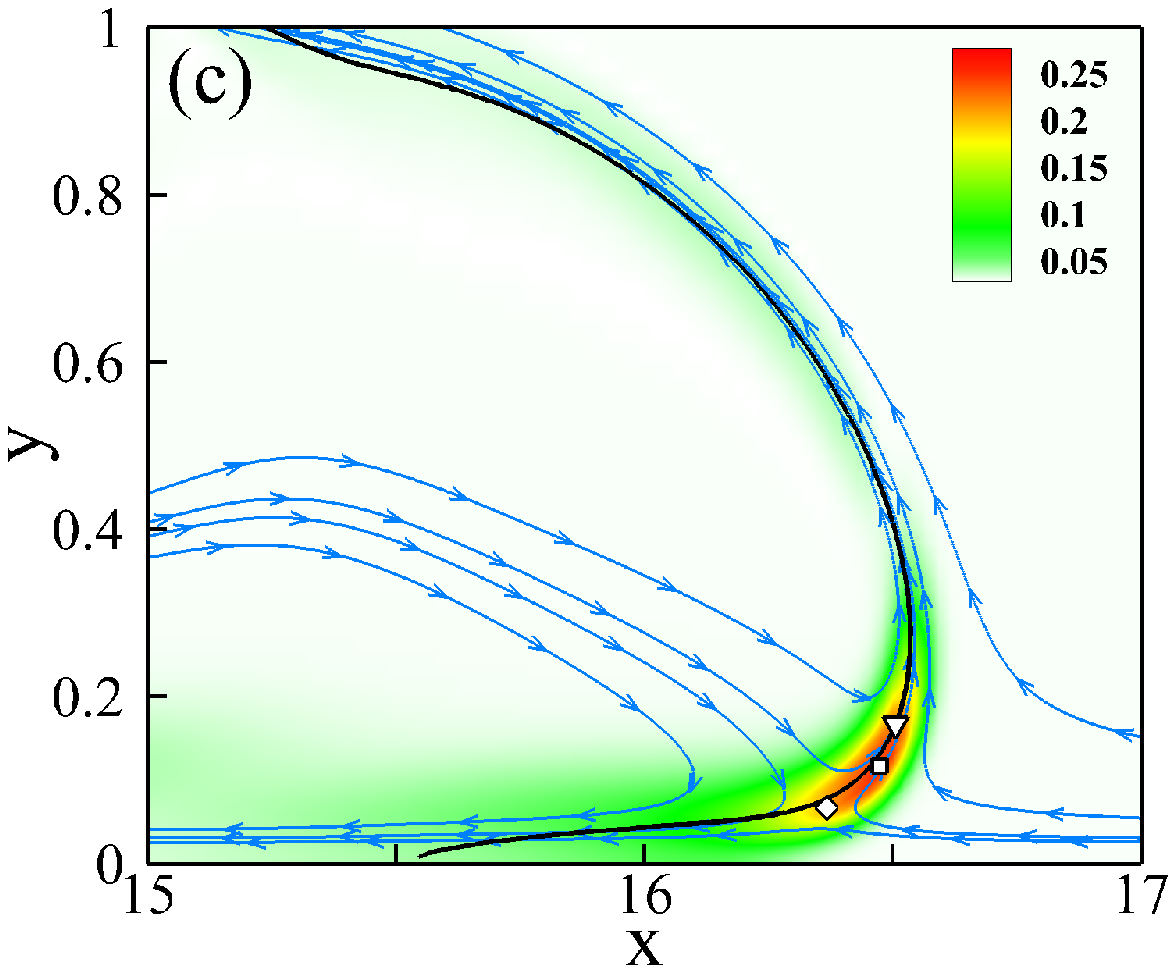}
\includegraphics[width=1.68in]{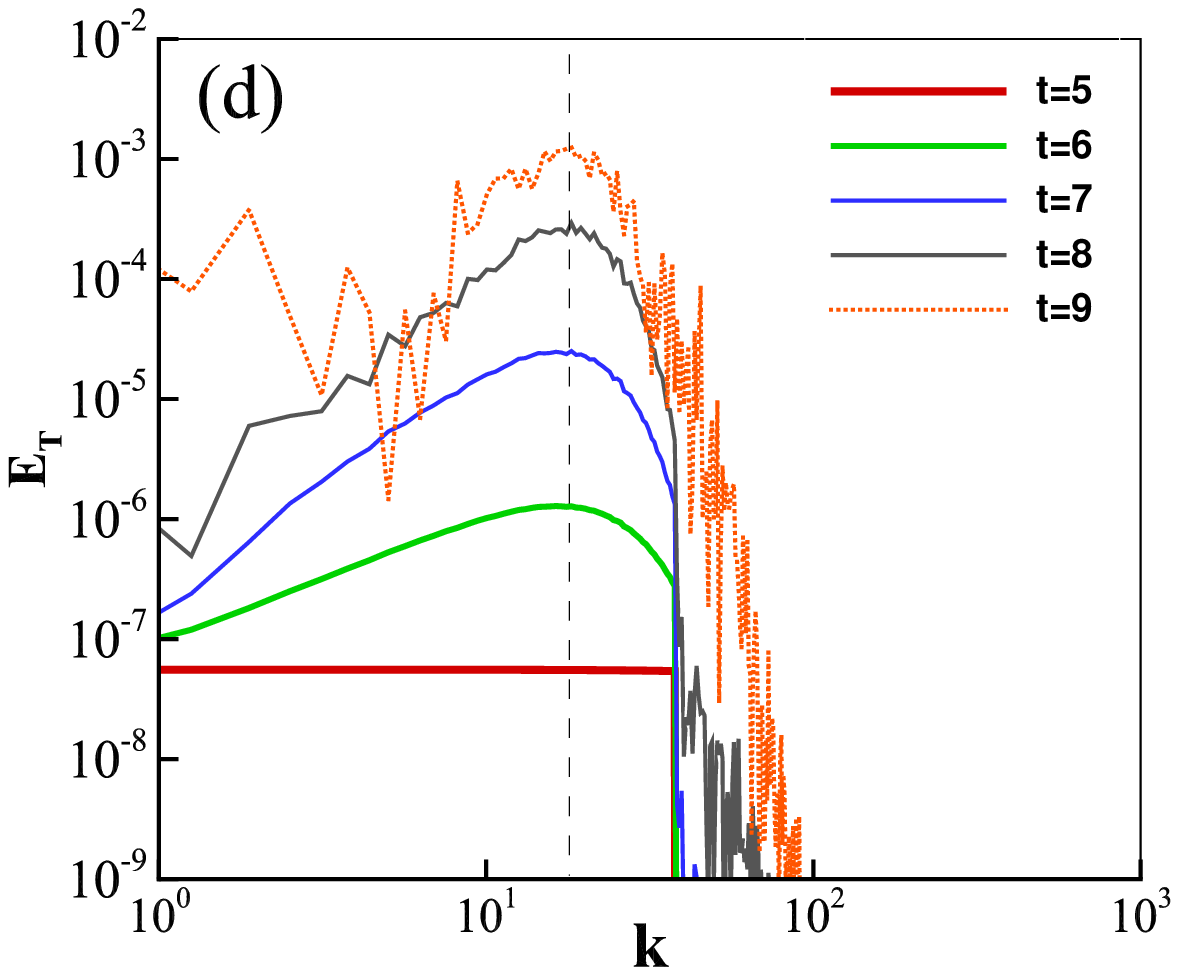}
\end{minipage}
\caption{The iso-surfaces of the dimensionless temperature $T=0.5$ at $t=8$ and 10 are shown in (a) and (b), respectively. The temperature perturbation amplitude $T^*_{max}$ at $t=8$ is shown in (c), where $T^*(x,y,t)_{max}=[T(x,y,z,t)-\bar {T}(x,y,t)]_{max}$ and the spanwise mean temperature  $\bar{T}=\frac{1}{Lz}\int_{0}^{Lz}T(x,y,z,t)dz$. The black solid line indicates $\bar{T}=0.5$ and the empty square, diamond, and triangle symbols represent the positions of the maximum $T^*_{max}$, the stagnation point of the spanwise mean flow, and the maximum curvature of the interface with $\bar{T}=0.5$, respectively. Some streamlines are drawn as well to show the mean flow field in the translational frame moving with the front.  (d) The energy spectral density $E_T$ of the temperature perturbation $T^*$ along the spanwise line where $T^*_{max}$ reaches its maximum at the front region. The vertical dash line indicates the dominating spanwise wavenumber, where the spectral density reaches the maximum. In the simulation, Gr$=10^6$, Pr=1, and the random disturbances are introduced at $t=5$. \label{fig:Graph2}}
\end{figure}

 It is shown in Fig. \ref{fig:Graph2}(a) that the spanwise modulation of the current front appears soon after the introduction of random disturbances, and develops quickly to the lobe and cleft structures [Fig. 2(b)]. The temperature perturbation amplitude $T^*_{max}$, representing the density amplitude, is shown in Fig. 2(c). During the slumping phase, the foremost part of the front in a translational frame moving with the front velocity is almost stationary. The strongest perturbation indicated by the empty square is very close to the interface of $\bar{T}=0.5$ and between the stagnation point of the mean flow and the maximum curvature of the mean interface averaged in the spanwise direction, agreeing with the previous observation  that the maximum amplitude of the most unstable mode prevails between the stagnation point and the nose \cite{Hartel2000b}. According to the previous linear stability analyses, the growth rate or amplification rate is a continuous function of the wavenumber and has a single maximum at the dominating wavenumber $k_{max}$ for the unstable modes (Fig. 5 of \cite{Hartel2000b}). Therefore, at the linear evolution stage where the perturbation amplitudes are small, the energy spectra of temperature perturbations caused by random disturbances with the same initial amplitude $E_T$, which satisfies $\frac{1}{L_z}\int^{L_z}_0 T^{*2}dz=\int E_T (k)dk$, should be a smooth curve and has only one peak at $k_{max}$. Consequently, the original dominating spanwise wavenumber can be determined strictly by finding the wavenumber with the maximum spectral density at the initial stage of the instability. As illustrated in Fig. 2(d), the dominating wavenumber $k_{max}=18.2$.  $n_{dist}=60$ corresponds to a wavenumber of $37.7>k_{max}$, and hence the wavenumber range of the introduced disturbances is wide enough to cover the dominating wavenumber. It is also checked that the dominating wavenumbers obtained with $n_{dist}=60$ and 100 are the same. During the nonlinear evolution stage, some lobes grow larger than their neighbors and some small lobes may disappear due to the cleft merging, and hence the spectral curve becomes irregular and has some peaks at low wavenumber range as the curves with $t>7$ shown in Fig. 2(d).

 In order to clarify the contribution of the velocity field to the lobe-and-cleft formation, a new reference frame is built as shown in Fig. 1(a), where $\bar{x}$, $\bar{y}$ and $\bar{z}$ are in the tangent, normal and spanwise directions of the interface of $\bar{T}=0.5$, respectively. $\bar{y}$ deviates from $y$-direction with an angle of $\theta$,  and the origin of the $\bar{x}\bar{y}\bar{z}$ frame has a distance of $\bar{h}$ from the bottom wall in the $\bar{y}$ direction. Now we study the Rayleigh-Taylor instability (RTI) only based on the temperature field in the $\bar{y}-\bar{z}$ plane, where the acceleration is $\bar{g}=g\cos\theta$. Because of the diffusivity the density interface has a finite thickness, whose stabilising effect on the RTI can be estimated by an effective Atwood number $\frac{A}{1+AkL}$ \cite{Lelevier1955,Cherfils1996,Mikaelian1997} or an effective Grashof number Gr$_{e}=\frac{Gr}{1+AkL}$ and the temperature gradient length $L=(\bar{T}/\frac{d\bar{T}}{d\bar{y}})_{\bar{T}=0.5}$.

 Assuming that the normal mode is in the form of $f'=f_{1}(\bar{y})exp[ik\bar{z}+\gamma t]$, where $f'$ represents the disturbing $\bar{y}$-direction velocity $v'$, temperature $T'$, and pressure $p'$ respectively, and substituting them into the linearized governing equations in the $\bar{y}-\bar{z}$ plane, we have
\begin{eqnarray}
\label{eq:eq-4}
 &&(\gamma+\frac{1}{\sqrt{Gr}Pr}D)T_{1}=-v_{1}\frac{dT_{0}}{d \bar{y}},\\
  &&(\gamma+\frac{1}{\sqrt{Gr}Pr}D)(\gamma+\frac{1}{\sqrt{Gr}}D)Dv_{1}= -\frac{Gr_e}{Gr}\cos(\theta)k^2v_{1}\frac{d T_{0}}{d\bar{y}}  ,\ \ \ \\
  &&D=k^2-\frac{d^2}{d \bar{y}^2},\nonumber
\end{eqnarray}
where $T_0$ is the temperature field obtained in simulations when the current flow is still two dimensional. These equations can be solved with a method similar as used in \cite{Mikaelian1996}. The eigenfunctions $T_{1}$ and $v_1$ can be expressed as
\begin{equation}
\label{eq:eq-6}
  T_{1}=\begin{Bmatrix} A_{11}e^{\alpha_{1}\bar{y}}+A_{12}e^{-\alpha_{1} \bar{y}},-\bar{h}\leq \bar{y}\leq0\\ A_{21}e^{-\alpha_{1}\bar{y}}+A_{22}e^{\alpha_{1} \bar{y}},0< \bar{y} \end{Bmatrix},
\end{equation}
\begin{equation}
\label{eq:eq-7}
  \bar{v}_{1}=\frac{1}{\sqrt{Gr}}\begin{Bmatrix} B_{11}e^{k\bar{y}}+B_{12}e^{-k\bar{y}}+C_{11}e^{\alpha_{2}\bar{y}}+C_{12}e^{-\alpha_{2}\bar{y}}\\
  +A_{11}^{'}e^{\alpha_{1}\bar{y}}+A_{12}^{'}e^{-\alpha_{1}\bar{y}},-\bar{h}\leq \bar{y}\leq0\\
  B_{21}e^{-k\bar{y}}+B_{22}e^{k\bar{y}}+C_{21}e^{-\alpha_{2}\bar{y}}+C_{22}e^{\alpha_{2}\bar{y}}\\
  +A_{21}^{'}e^{-\alpha_{1}\bar{y}}+A_{22}^{'}e^{\alpha_{1}\bar{y}},0< \bar{y}  \end{Bmatrix},
\end{equation}
where $A_{ij}^{'}=\frac{Gr_e\cos(\theta)k^2}{GrPr(Pr-1)\gamma^2}A_{ij}$ with $i$ and $j=1,2$, $\alpha_{1}=\sqrt{k^2+Pr\sqrt{Gr}\gamma}$, and $\alpha_{2}=\sqrt{k^2+\sqrt{Gr}\gamma}$.

It is known that the influence of the unstable RT mode is limited in the range about one wavelength from the interface \cite{Drazin02}. Since the dominating spanwise wavelength is smaller than the height of the current front as shown in Figs. 2(a) and 2(b), the unstable modes decay exponentially in the dense fluid $(\bar{y}>0)$ and vanish at the bottom of the light fluid where $\bar{y}=-\bar{h}$. By applying the boundary conditions and the continue conditions for velocity, stress, and temperature at the interface, Eqs. (4)-(7)  can be written in a form of $MV=0$, where $M$ is an $12\times12$ matrix and $V$ is a vector composed by the twelve constants, $A_{ij}$, $B_{ij}$, and $C_{ij}$. A nontrivial solution requires that the determinant of $M$ equals zero, and hence the dispersion relation $D(Gr, Pr, \theta , L,  k, \gamma )=0$ is obtained. For details of the solving method, we refer to \cite{Mikaelian1996}.

\begin{figure}
\centering
\begin{minipage}[b]{0.50\textwidth}
\centering
\includegraphics[width=1.72in]{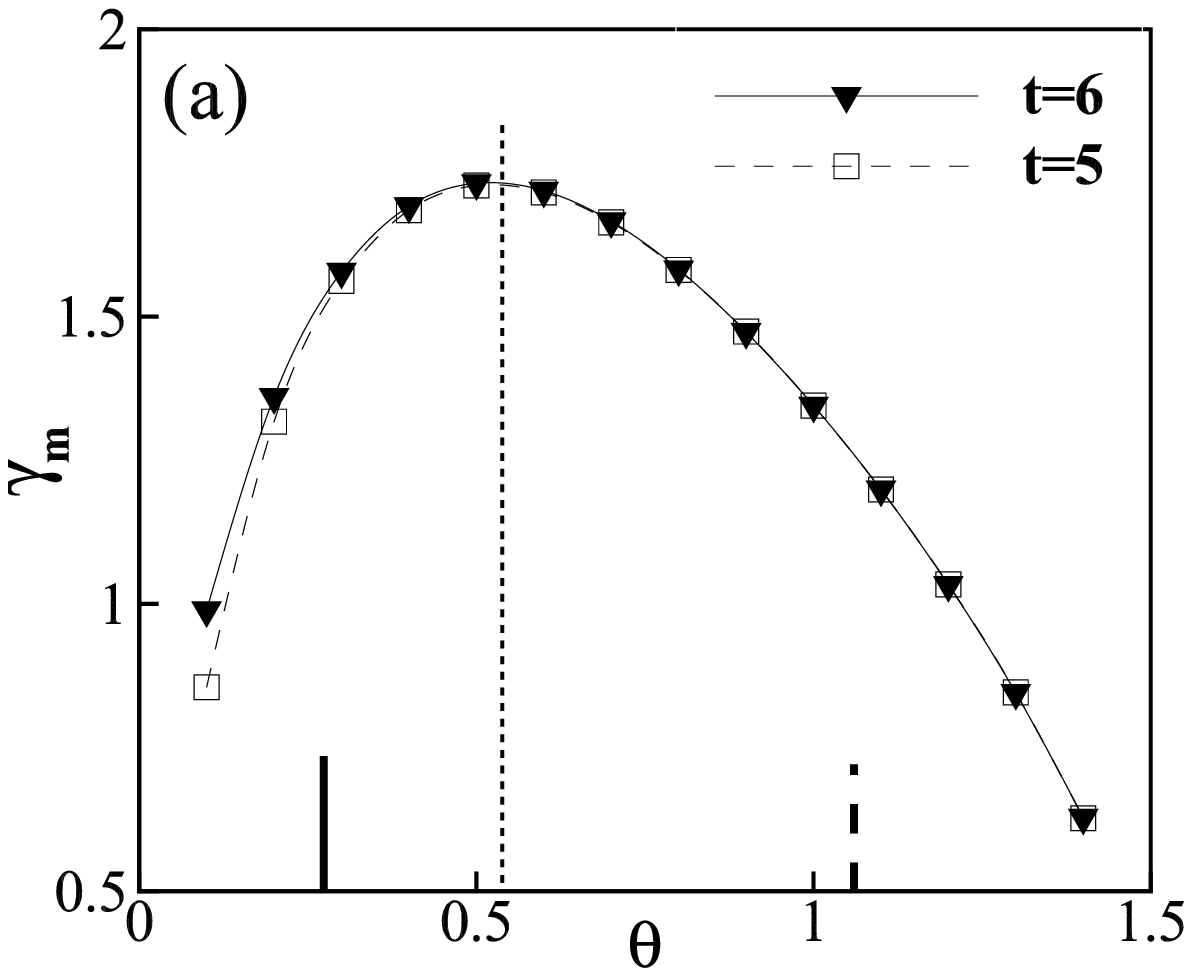}
\includegraphics[width=1.72in]{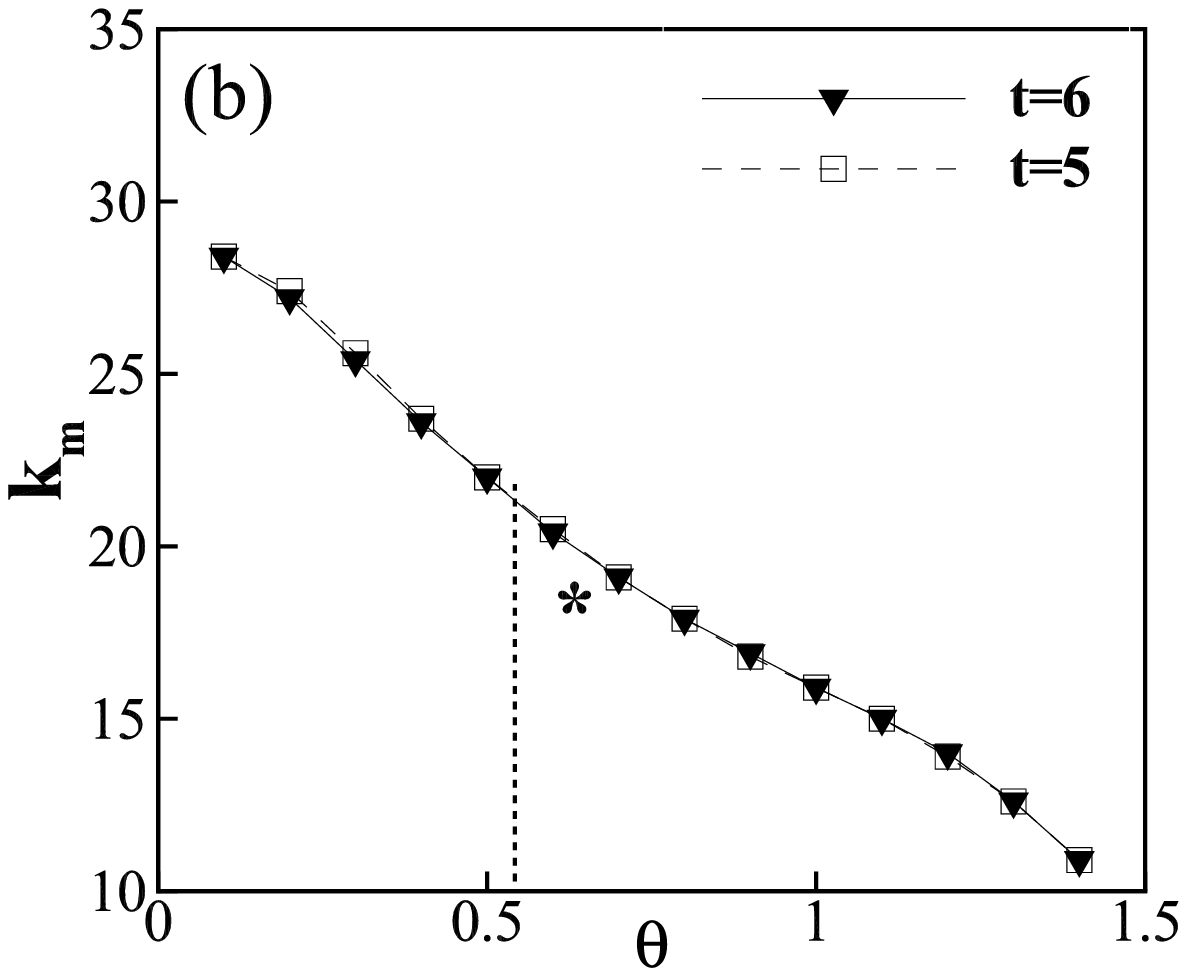}
\end{minipage}
\caption{(a) The maximum growth rate $\gamma_{m}$ and (b) the corresponding wavenumber $k_{m}$ calculated from the local dispersion relation as functions of $\theta$ for Pr=1 and Gr$=10^6$. The short solid and dash lines in (a) represent the positions of the stagnation point and the maximum curvature, respectively. The symbol $\star$ in (b) indicates the value of the strongest perturbation obtained from simulations.  \label{fig3}}
\end{figure}

\begin{figure}
\centering
\begin{minipage}[b]{0.50\textwidth}
\centering
\includegraphics[width=2.6in]{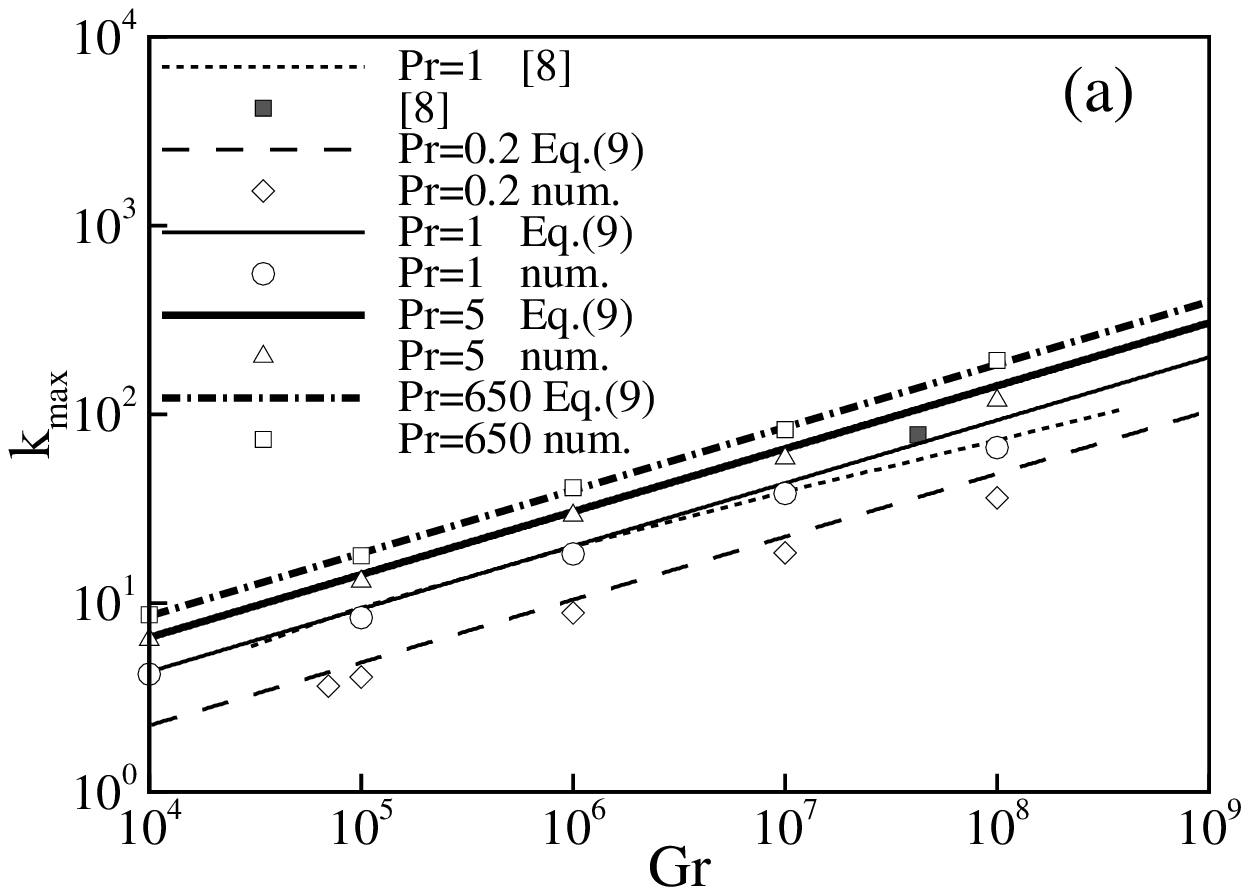}
\includegraphics[width=2.6in]{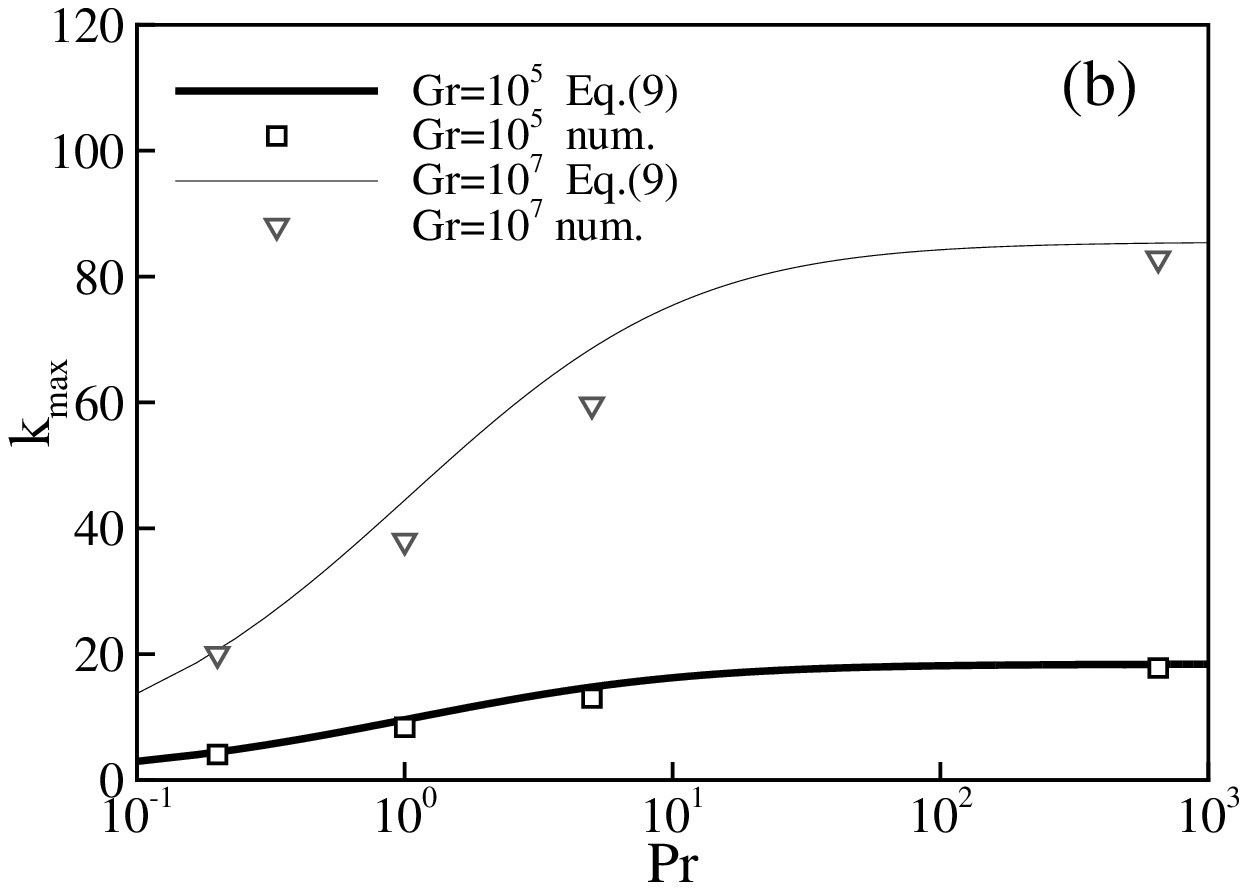}
\end{minipage}
\caption{ The original dominating wavenumber obtained in simulations as a function of (a) the Grashof number and (b) the Prandtl number. The theoretical curves of the semi-infinite RTI model [Eq. (9)] are shown in (a) and (b) as well for comparison. The filled square in (a) represents the estimated mean wavenumber \cite{Hartel2000b} based on  Simpson's salt and fresh water experiments \cite{Simpson1972}, where the equivalent of $Pr$ (Schmidt number) is about 650.  \label{fig4}}
\end{figure}

The maximum growth rate $\gamma_{m}$ and the corresponding spanwise wavenumber $k_{m}$ are calculated according to the local dispersion relation for given positions ($\theta$) along the interface. It is shown in Fig. \ref{fig3}(a) that the maximum of $\gamma_{m}$, which is referred as $\gamma_{max}$,  lies at $\theta =0.54$ for Pr=1 and Gr=$10^6$. Stability analyses were carried out based on the temperature fields obtained at different instants (t = 5 and t=6) respectively, and $\gamma_{m}$ and $k_{m}$ curves almost coincide with each other. As shown in Fig. 3(b), $k_{max}$ corresponding to the maximum of $\gamma_{m}$ is 21.3,  which is close to the dominating spanwise wavenumber 18.2 obtained from the numerical simulations [Fig. 2(c) and 2(d)], indicating that the origin of the lobe and cleft is determined mainly by the Rayleigh-Taylor instability at the front lower interface. This result is surprising at its first glance because the present stability analysis is solely based on the density field, while the previous gravitational stability analyses \cite{Hartel2000b} required detailed information of both the velocity field and the density field of the current front. The underlying rationality is based on two features of the mean flow. First, the mean flow around the position of the strongest perturbation in the translational frame is weak, because it is close to the stagnation point, whose mean velocity is zero. It is noted that the front velocity field without perturbations is two-dimensional and has no spanwise velocity. Second, as shown by the streamlines in Fig. 2(c), the mean flow around the position of the strongest perturbation is nearly parallel to the interface or in the $\bar{x}$ direction, and then has little contribution to the RTI in the $\bar{y}\bar{z}$ plane.

The acceleration and the interface thickness can change evidently the growth rate of RT unstable mode but not the dominating wavenumber, which is mainly determined by the momentum and the density diffusion effects \cite{Duff1962}.  The agreement between the simulation and the RTI model shown in Fig. 3(b) encourages us to consider a further simplification, where the exact density and velocity fields of the current front are not required. Following the previous studies \cite{Duff1962,Bellman1954,Renaud1997}, the dispersion relation governed by Eqs.(4)-(7) can be simplified for the semi-infinite fluids with a planar and infinitely thin interface,  and the approximate but analytical solution is solved as
\begin{equation}
\label{eq:eq-9}
\gamma=\sqrt{\frac{k}{2}+\frac{k^4}{Gr}}-\frac{1}{\sqrt{Gr}}(1+\frac{1}{Pr})k^2.
\end{equation}
The most unstable mode corresponds to $d\gamma/dk=0$, and it is easy to obtain that
\begin{equation}
\label{eq:eq-11}
k_{max}=({\frac{Gr}{2}})^\frac{1}{3}({\frac{1}{[2(1+\frac{1}{Pr})+\sqrt{4(1+\frac{1}{Pr})^2-3}]^2-1}})^{\frac{1}{3}},
\end{equation}
indicating that the original dominating spanwise wavenumber should scale as Gr$^{1/3}$. For Gr$=10^6$ and Pr=1, Eq. (9) suggests $k_{max}=20.1$, which is close to the simulation value 18.2. By analyzing the simulation data with the same methods as discussed in Fig. 2 and Fig. 3, the dominating wavenumbers are obtained and agree well with the theoretical predictions for different Grs and Prs as shown in Fig. \ref{fig4}(a), indicating clearly the 1/3 scaling law and confirming that RTI is the dominating onset mechanism for lobes and clefts. This mechanism explains why $k_{max}$ was found numerically not sensitive to the length of the computational domain $L_x$: Short domain or small $L_x$ may affect the flow topology and the front speed due to the periodic boundary condition, but $Pr$ and $Gr$ remain the same and hence the wavenumber determined by RTI does not change.

When Pr$\rightarrow \infty$, it is easy to obtain the asymptotic value of $k_{max}=Gr^{1/3}/2^{4/3}$ from Eq.(9). For Boussinesq fluids, such a Pr dependence is reasonable because small Prandtl number corresponds to strong density diffusion, which suppresses high wavenumber components and lower the dominant wavenumber. As shown in Fig. 4(b), $k_{max}$ decreases with the decrease of Pr and has a strong dependence on Pr for Pr$<50$ as predicted by the semi-infinite RTI model. In fact, for the cases with small $Pr$ and $Gr$, the strong density and momentum diffusion effects thicken the density interface and shorten the light fluid layer beneath the dense one at the current front, and hence undermine the prerequisite of RTI. According to the simulations for $Pr$=0.2, the current front becomes stable to spanwise perturbations when $Gr$ is as low as $10^4$.

The spanwise wavenumber of the most unstable mode obtained previously by analyzing a rectangular domain \cite{Hartel2000b} are shown as a dotted curve in Fig. 4(a), agreeing well with the present simulations. When lobes are clearly recognized, the nonlinear growth of lobes and cleft merging have occurred, leading to notable growth of low wavenumber components as shown in Fig. 2(d). Consequently, the mean wavenumber is smaller than the original dominating value, especially when it is estimated by counting the lobe's number. For example, the estimated spanwise wavenumber of Simpson's experiments \cite{Simpson1972} is shown in Fig. 4(a) as the filled square \cite{Hartel2000b}, which is lower than the simulation data.

\section{Conclusions}
 Though the velocity field and the density field are coupled for the gravity currents, the present simulations and stability analyses show that the local Rayleigh-Taylor instability at the density interface can determine the position and the original spanwise wavenumber of the strongest perturbations, illustrating that the density stratification is the crucial factor dominating the onset of lobes and clefts. Without any detailed information of the flow field but only the fluids' properties ($Gr$ and $Pr$), the semi-infinite RTI model predicts successfully the original dominating spanwise wavenumber of the current front and the 1/3 scaling law, indicating that RTI is the original formation mechanism for lobes and clefts. In addition, it is shown theoretically and numerically that the Prandtl number has substantial effect on the spanwise wavenumber selection.

The authors thank Paul Linden for enlightening discussions on the current flows. Simulations code SIMSON from KTH and the help from Philipp Schlatter, Luca Brandt and Dan Henningson are gratefully acknowledged. The simulations were performed on TianHe-1(A) and this work has been supported by the National Natural Science Foundation of China (Grants No. 91752203, No. 11490553, and No.11521091).


\end{document}